\documentclass[nofootinbib,twocolumn]{revtex4}

\usepackage{amsmath}
\usepackage{amsthm}
\usepackage{amsfonts}
\usepackage{amssymb}
\usepackage{dsfont}
\usepackage{color}
\usepackage{mathrsfs}

\usepackage{graphicx}
\usepackage{epstopdf}
\usepackage{slashed}
\newcommand{\ie}{{\emph{i.e.}}~}
\newcommand{\eg}{{\emph{e.g.}}~}

\theoremstyle{plain}

\newcommand{\ee}{{\mathbf{e}}}

\newcommand{\BB}{\mathscr{B}}

\def\be{\begin{equation}}
\def\ee{\end{equation}}
\def\ba{\begin{align}}
\def\ea{\end{align}}
\def\trace{\textrm{Tr}}

\newcommand{\ket}[1]{\vert #1 \rangle}
\newcommand{\bra}[1]{\langle #1 \vert}
\newcommand{\proj}[1]{\vert #1\rangle\!\langle#1 \vert}

\newcommand{\dd}{\textrm{d}} 

\newcommand{\iu}{\mathrm{i}}
\newcommand{\e}{\mathrm{e}}

\newcommand{\hdet}{\hat{H}_{UDW}}

 \makeindex

\begin{document}
\title{Quenches and lattice simulators for particle creation} 
\author{Francesco Caravelli${}^{1,2}$}
\email{francesco.caravelli@gmail.com}
% \author{Alioscia Hamma${}^{1}$}
% \email{hamma@perimeterinstitute.ca}
\author{Fotini Markopoulou${}^3$}
\email{fotinimk@gmail.com}
\author{Arnau Riera${}^{3,4,5}$}
\email{arnauriera@gmail.com}
\author{Lorenzo Sindoni${}^{4}$}
\email{sindoni@aei.mpg.de}

\affiliation{
${}^1$Santa Fe Institute for Complex Systems,
   1399 Hyde Park Road,
   Santa Fe, NM 87501, US\\
and\\
${}^2$OCIAM Mathematical Institute, University of Oxford, 24-29 St. Giles,
Oxford, OX1 3LB, UK,\\and\\
${}^3$ Perimeter Institute for Theoretical Physics, \\
Waterloo, Ontario N2L 2Y5
Canada, \\and\\
${}^4$ Max Planck Institute for Gravitational Physics, Albert Einstein Institute,\\
Am M\"uhlenberg 1, Golm, D-14476 Golm, Germany\\
and\\
${}^5$Dahlem Center for Complex Quantum Systems, Freie Universit{\"a}t Berlin, 14195 Berlin, Germany
}

\begin{abstract}
In this paper we propose a framework for simulating approximate thermal particle production in 
condensed matter systems. The procedure we describe can be realized by means of a quantum quench of 
a parameter in the model.

In order to support this claim, we study quadratic fermionic systems
in one and two dimensions by means of analytical and numerical techniques. In particular, we are able to show that a class of 
observables associated to Unruh--de Witt detectors
are very relevant for this type of setup and that exhibit approximate thermalization. 

\end{abstract}
 \maketitle
\section{Introduction}

The relative weakness of the gravitational field makes difficult any direct test of some of the 
phenomena that should characterize quantum field theory (QFT) in curved spacetimes, namely various
instances of particle creation from the quantum vacuum. For the same reason, the regime in
which quantum gravitational phenomena deviate significantly from the predictions of Einstein
gravity (viewed as the effective field theory description of a full-fledged quantum gravity theory)
seems to be still out of reach. 
 Therefore simulators, \ie models offering analogies with gravitational
systems,  could help us to develop concepts, methods and insights for the 
investigation
of the relationship between quantum mechanics and the geometry of spacetime.

Since the seminal work of Unruh \cite{Unruh}, who showed how four dimensional Lorentzian geometries can be 
simulated in hydrodynamical systems, the area of analogue models has flourished \cite{AnalogueReview}, providing 
nowadays a rather large number of (classical and quantum) models based on condensed matter systems,
which are able to simulate QFT in curved spaces.
Furthermore, we have by now numerical and experimental results \cite{Weinfurtner:2008if,Carusotto,stimulated,HawkingFilaments,HorstmannLett,HorstmannLong,Rubino:2011zq} which allow us to discuss
concretely the validity of the theoretical predictions for phenomena
like Hawking radiation and cosmological particle creation\footnote{The crucial point is the
fact that these phenomena do not need Einstein's equations for the metric to be satisfied. They are in a sense kinematical, as far as the metric tensor is
concerned.}.

 Most of the analogues are based on continuous approximations, hence applying methods of
continuuum field theories. However, this is not the only possibility that we have.
It has been already proposed to use lattice systems to model new interesting 
situations. For instance, the use of optical lattices has been advocated in \cite{HuRey1} and 
\cite{SchuetzSzpak1,SchuetzSzpak2} in order to explore regimes in which conventional continuum
mean field descriptions fail or in which the quantum vacuum is explored in the strong external
field limit, regimes difficult to achieve and to control in other quantum systems like 
BECs\footnote{For instance, the curved acoustic geometry of BECs is related to the velocity of the condensate flow, 
its density, and derivatives. To achieve strong gravity effects, then, requires manipulations of these quantities
without destroying the condensate regime.}.

Another motivation for discrete analogues comes from the number of discrete Quantum Gravity proposals.
In particular, the possibility that at a microscopic level continuum spacetime 
might be
replaced by a completely combinatorial structure in the shape of a random graph (variously decorated with geometric data) is a widespread
idea \cite{Approaches}.
Despite the variety of proposals, 
the manipulation of random graphs (or random complexes) within a 
reparametrization invariant framework presents difficulties which are so far not
completely under control, especially if we try to obtain a quantitative explanation of the emergence 
of classical continuum spacetime and its dynamics. 

In fact, in \cite{QGp1,QGp2,QGp3,QGp4} it has been proposed to address a simpler problem, that is the emergence 
of space alone, as the outcome of the collective dynamics
a certain specific class of (quantum) Hamiltonians for (quantum) graphs: Quantum Graphity.
Since in such a framework there is a
preferred foliation related to the presence of a preferred notion of time 
(here radically different from space),
the inclusion of Lorentz invariance (and of the full four dimensional diffeomorphism invariance) is
problematic\footnote{It is not obvious that 
it is impossible to reconcile these models
with a gauge fixed version of Lorentz and diffeomorphism invariant models. It is
likely that in the continuum limit, these would be described by effective theories of
the class of theories recently introduced by Ho\v{r}ava \cite{Horava} and extensively discussed in the literature, as it happens in Causal Dynamical Triangulations \cite{cdthl}.}. Nonetheless, they represent
ideal candidates to discuss in an explicit way at least some general features of the otherwise
vague idea that space, time and geometry are emerging from pregeometric degrees
of freedom.
For such a purpose, different models have been introduced so far, with the objective of
elucidating the way in which the dimensionality of the graphs is dynamically controlled, as well as
the properties of the phase transition from the disordered, pregeometric phase to the ordered, geometric one  \cite{QGp4,Conrady,Chen}.
Possible cosmological effects due to their fundamental discrete structures have been considered recently in \cite{Quach}.

As we shall argue in detail in the rest of the paper, these systems might still offer interesting insights as far as quantum
field theory in curved spaces is concerned. 
%{\sf Sentence deleted.} 
In particular, we will show how these models can be used to simulate particle creation
phenomena (either in a black hole-like configuration, like in \cite{BlackOnions}, or in a cosmological
setting) by analogy with Unruh effect, tying it to quench experiments, which are indeed
the natural experimental setup.

Concretely, we will work with a Fermi--Hubbard model on a fixed lattice with an additional term in the Hamiltonian, which does 
not conserve the particle number. 
%\sout{The same Hamiltonian describes the physics of the Josephson junction.} 
A similar model appeared already in the literature of Lieb--Robinson bounds and has been experimentally 
probed \cite{optlattnat}. 

The model which we consider has been initially derived from Quantum Graphity \cite{Hamma}, as an
effective description in which the graph is essentially frozen in a given configuration. The derivation is explained in Section IV in the 
section dedicated to Quantum Graphity.
The 
application of our setup to quantum graphity's trapped surfaces, for the sake of clarity, is at the very end of the paper,
as in principle this setup could apply to other physical models.

The paper is organized as follows. Section II is devoted to the detailed explanation
of the model and of the calculations that will be presented, in particular with respect to
the notion of particle and detectors. 
In section III we first show the numerical results on the thermality of the detected particles,
for various 2-dimensional square-shaped lattices and for the 1-dimensional ring. The latter is treated analytically in full detail. 
A subsection is devoted to the clarification of the results from the point of view 
of quantum quenches. Section IV explains the application to quantum graphity's trapped surfaces,
with emphasis on the subtle but important differences with the case of true black holes.
Some final remarks conclude the paper.

\section{The analogy}
\subsection{Time independent Hamiltonians: detectors}

Contrary to what could happen in a general relativistic context, in non-relativistic quantum mechanics, the notion of ground state
(vacuum) and excited states (particles) is globally defined (see, \eg, \cite{BirrellDavies}). 
Moreover, if we consider static configurations (\ie time-independent Hamiltonians), any energy eigenstate 
is preserved by the unitary time evolution and, thus, particle creation effects cannot take place.

In order to overcome this obstacle and make radiation effects possible, 
it is necessary that the notion of particle that the detector measures does not correspond to
the excitations of the Hamiltonian. 
This is not so unnatural since 
detectors are local objects, and therefore, 
there is no reason to think that they are be able to capture the precise structure of the eigenstates of the global Hamiltonian.

In particular, we need to introduce a notion of particle (with an associated momentum $k$ and energy $\epsilon(k)$)
given in terms of some ladder operators $\eta_k$. 
Once the notion of particle that the detector measures is established, we
can determine the momentum distribution (number of particles with momentum $k$) of the ground state of the system 
\begin{equation}
n(k) = \bra{GS} \eta_k^\dagger \eta_k \ket{GS} ,
\label{eq:momentum-distribution}
\end{equation}
and see whether it follows a thermal distribution.

The notion of particle $\eta_k$, together with its associated momentum $k$ and energy $\epsilon(k)$,
define a test Hamiltonian $H_{UDW}=\sum_k \epsilon(k) \eta_k^\dagger \eta_k$. 
The momentum distribution \eqref{eq:momentum-distribution} that the detector measures is 
given by the overlap between the ground state of the system and the eigenstates states of the test Hamiltonian $H_{UDW}$.

\subsection{Time dependent Hamiltonians: quenches}
Another possibility for simulating particle creation phenomena is to consider 
time dependent Hamiltonians.
If some parameter of the Hamiltonian changes fast enough, the adiabatic approximation breaks down
and the system leaves its ground state.
The populations of the excited states become non trivial and hence radiation is produced.

An example of such a process is a time dependent tunnelling amplitude between the sites of the lattice.
This can be interpreted as a time dependent
scale factor of the spatial geometry, giving the possibility to study, along the very same lines 
described for a black hole configuration, analogue cosmological particle production phenomena.

A particularly interesting case of a time dependent Hamiltonian is the quench setting.
A quench experiment consists of preparing a quantum system in the ground state of a certain Hamiltonian,
and, suddenly, changing the Hamiltonian such that the quantum state does not correspond to an eigenstate of the
new Hamiltonian anymore. The system, then, is out of equilibrium an evolves non-trivially in time.
The goal of quench experiments is to study the time evolution of systems and in particular, its possible 
relaxation to equilibrium.

Let us notice that the process of equilibration after a quantum quench for the whole system is, strictly speaking, never possible. 
This is due to the fact that the system evolves 
according to a unitary evolution. Nevertheless, what we mean by equilibration is the relaxation
of the expectation values of a certain family of observables to constant quantities. 
The question whether the equilibrium state after a quench corresponds or not to
a thermal state has been studied in several works  \cite{calabrfag,foinifluctdiss,essevfag,calabrese06,calabrese07}.

If the Hamiltonian is quenched to the  Hamiltonian that defines the notion of particle in the detector setting $H_{UDW}$, 
then the momentum distribution of the system at any time after the quench is identical to the momentum distribution
measured by the detector in the time independent case, and hence, both settings become equivalent.
This is due to the fact that the observables $\eta_k^\dagger \eta_k$ commutes with the Hamiltonians considered 
before and after the quench.

The equivalence between the two settings suggests
the use of quench experiments to simulate particle production also in the detector framework.
Furthermore, the fact that the momentum distribution does not change along the evolution dictated by $\hdet$
allows to measure it at the most convenient time.

%%%%%%%%%%%%%%%%%%%%%%%%%%%%%%%%%%
%%%%%%%%%%%%%%%%%%%%%%%%%%%%%%%%%%
%%%%%%%%%%%%%%%%%%%%%%%%%%%%%%%%%%

\subsection{The model}

The Hamiltonian we will consider in the present paper is the following Fermi--Hubbard model:

\be
H(\Gamma,\lambda)=-J\sum_{i,j=1}^{N_v} A_{ij}^{(\Gamma)}a^\dagger_ia_j+\frac{\lambda }{2}\sum_{i,j=1}^{N_v} B_{ij}^{(\Gamma)}
\left(a^\dagger_ia^\dagger_j+\textrm{h.c.}\right),
\label{eq:full-hamiltonian}
\ee
where $a_i$,$a^\dagger_i$ are annihilation/creation fermionic operators that annihilate/create a particle in the vertex $i$
of the background graph $\Gamma$. 
The matrices $A_{ij}^{(\Gamma)}$ and $B_{ij}^{(\Gamma)}$ are, respectively, the adjacency matrix of $\Gamma$
and its antisymmetrized form\footnote{The antisymmetrization is arbitrary up to a certain extent. In fact, $B_{ij}=\sigma(i,j) A_{ij}$, with $\sigma(i,j)$ being any function such that $\sigma(i,j)=-\sigma(j,i)$ and $|\sigma(i,j)|=1$ $\forall i,j$.
However this does not change the properties of the system as long the graph is chosen properly.}. 
The sum runs over all the $N_v$ vertices of the graph $\Gamma$.
The coupling $J$ is the tunneling of the particles between two connected sites and $\lambda$ controls the strength of the Hamiltonian terms
that do not conserve the number of particles.

In quantum graphity, the role of curvature in a continuous space-manifold is played by the connectivity of a dynamic graph, while the backreaction of matter on geometry, in
graphity models, is controlled usually
 by a term in the Hamiltonian that annihilates particles and creates
links of the graph\footnote{ 
In order that the Hamiltonian is Hermitian, the opposite process is also required.}, with the idea that the larger is the number of particles
in a region, the larger is the effect on the connectivity of the curvature. 

We derived this model as an effective description of Quantum Graphity in the limit of 
a very dilute matter content and a weak backreaction term between the particles and the graph (see Sec.~IV). 
This violation of the number of particles related to the connectivity of the graph
is produced in Hamiltonian \eqref{eq:full-hamiltonian} by the second term proportional to $\lambda$. 
In a more rigorous way, Hamiltonian \eqref{eq:full-hamiltonian} is exactly obtained from the quantum graphity Hamiltonian
after freezing the evolution of the graph and assuming that this is in a superpositon state (see Sec.~IV for details).

Let us mention that a 
similar effective model has been studied both theoretically and 
experimentally in the context of optical lattices \cite{optlattnat} in order to give experimental evidence to the Lieb--Robinson bounds. 
We thus expect that the results of our paper could be potentially tested experimentally, given that the experimenter should be able to measure
our observables.
  
\subsection{Detector and notion of particle}
In order to study particle creation effects, 
we need to specify the notion of particle (and its corresponding energy) that our detector will measure.
In a continuous flat space, we can define particles as plane waves with a well defined momentum $\vec k$ and energy $\hbar^2 k^2 /(2m)$. 
We have to generalize this idea of particle to graphs. 

The simplest graph one can consider is the discretization of the line or a ring lattice 
(in which we have under control irrelevant IR divergences). 
In the ring graph in particle, it is natural to define the annihilation 
operator of a particle with momentum $k$
as the discrete Fourier transform of the annihiliation operators in position $a_j$, 
\be
\eta_k=\frac{1}{\sqrt{L}}\sum_{j=0}^{L-1}\e^{\iu 2\pi j k /L} a_j \, , \nonumber
\ee
with $L$ the number of sites of the ring.
Notice that these are precisely the eigenmodes of the hopping Hamiltonian
\begin{equation}
H_{\textrm{Ring}}=-\sum_{j=0}^{L-1} (a_j^\dagger a_{j+1} +\textrm{h. c})
=\sum_{k=0}^{L-1} \epsilon(k) \eta_k^\dagger \eta_k \, ,\nonumber
\end{equation} 
where $\epsilon(k)=-2\cos(2\pi k / L)$ gives us the energy of the mode $k$.
In the continuum limit, both the wave function and the dispersion relation 
are recovered in their standard form.

Thus, when other lattice configurations are considered (torus, cylinder, etc.), it is natural to define 
particles
as the eigenmodes of the hopping Hamiltonian supported on a regular graph.
%the underlying regular lattice of the model \eqref{eq:connected-model}. 
This hopping Hamiltonian is written as
\begin{equation}
\hdet = \sum_{i,j=1}^{N_v} A_{ij}^{(\Gamma_0)} a_i^\dagger a_j=\sum_{k=1}^{N_v} \epsilon(k)\eta^\dagger_{k}\eta_{k},
\label{eq:H-detector}
\end{equation}
where $A_{ij}^{(\Gamma_0)}$ is the adjacency matrix of the regular graph $\Gamma_0$. 
The eigenmodes of $\hdet$, labelled by $k$, and with energy $\epsilon(k)$, define our notion
of particle. These are created and annihilated by the operators $\eta_k^\dagger$ and $\eta_k$, 
and are the excitation the detector measures.

Let us mention that while $\Gamma_0$ is a regular graph, in general,
the graph $\Gamma$ that defines Hamiltonian \eqref{eq:full-hamiltonian} does not have to, 
since it will have regions with
a higher or a lower connectivity. 
Roughly speaking, we can construct $\Gamma$ by taking the flat graph $\Gamma_0$,
and adding or removing links to/from it.

\subsection{Calculation}

The aim of our calculation is to probe how many particles (as the ones defined as in the previous section) 
will be present in the ground state of the Hamiltonian of (\ref{eq:full-hamiltonian}). 
Therefore,
we need to compute 
\begin{equation}
n(k) = \bra{GS} \eta^\dagger_{k}\eta_{k} \ket{GS}\, ,
\label{counts}
\end{equation} 
and check whether the ground state, with respect to this notion of particle,
appears as a thermal state, 
\ie~the momentum distribution \eqref{counts} follows a Fermi--Dirac distribution.

In order for the computation of 
\eqref{counts} to make sense, the Hilbert space of the full Hamiltonian and the one of the
detector have to have some overlap (in terms of unitary mappings from one to the other).
It is sufficient that the two different Hamiltonians are defined as operators on the same Hilbert space, 
\ie~the graphs have the same number of nodes.

The Hamiltonian \eqref{eq:full-hamiltonian} is a standard quadratic model, hence, it can be diagonalized as
\begin{equation}
H = \sum_{j=0}^{L-1} \omega(j) \psi_{j}^\dagger \psi_{j},
\end{equation}
by means of a Bogoliubov transformation of the fundamental particle operators, $a_{i}^\dagger,a_{j}$.
In turn, these are related by another Bogoliubov transformations to the operators
$\eta,\eta^\dagger$. Then, the operators $\eta, \eta^\dagger$ will be connected to the $\psi,\psi^\dagger$ by
the Bogoliubov transformation that is the composition of the Bogoliubov transformations that
relate  $\psi,\psi^\dagger$ to $a,a^\dagger$ and $a, a^\dagger$ to $\eta, \eta^\dagger$.
It can be written formally as
 \begin{equation}
\eta_{k} = \sum_{j=0}^{L-1} \left( \alpha_{kj} \psi_{j} + \beta_{kj} \psi^\dagger_{j} \right)\, ,
\end{equation}
where $\alpha_{kj}$ and $\beta_{kj}$ are the Bogoliubov coefficents.
Particle creation effects will be clearly related to them.

Notice that in general the ground state of these fermionic models is
not just the Fock vacuum of the quasiparticles (the eigenmodes of the given Hamiltonian),
but it is a Fermi sphere due to the appearance of some negative energy modes
in the spectrum (actually, this is the reason why we are using a Fermionic model, since a Bosonic
one would lead to instabilities). This implies that the momentum distribution
$n(k)$ will receive a contribution just by the very presence
of real quasiparticles in the Fermi sphere.

%%%%%%%%%%%%%%%%%%%%%%%%%%%%%%%%%%
%%%%%%%%%%%%%%%%%%%%%%%%%%%%%%%%%%
%%%%%%%%%%%%%%%%%%%%%%%%%%%%%%%%%%

\section{Analysis}
In this section we present numerical and analytical results which support the 
claim of having particle production, at least in the sense of detectors. 
The key ingredient is the mismatch between the eigenspaces of the Hamiltonian 
governing time evolution and the eigenspaces of the Hamiltonian with which we set up the
detector. 

\subsection{Numerical results}

We study here three different lattice geometries: a cylinder, a torus, and a ring. 
% In all cases, a completely connected region is added to the lattice and, for such a graph,
% the ground state of Hamiltonian \eqref{eq:connected-model} is determined.
We evaluate the populations $\langle \eta_k^\dagger \eta_k \rangle$
in the ground state of the eigenmodes of the Hamiltonian $\hdet$ that defines our notion
of particles, that is, the Hamiltonian with $\lambda=0$.
Finally, we check if the distribution of the occupations with respect to the energy is thermal.

\begin{figure}
\includegraphics[scale=0.5]{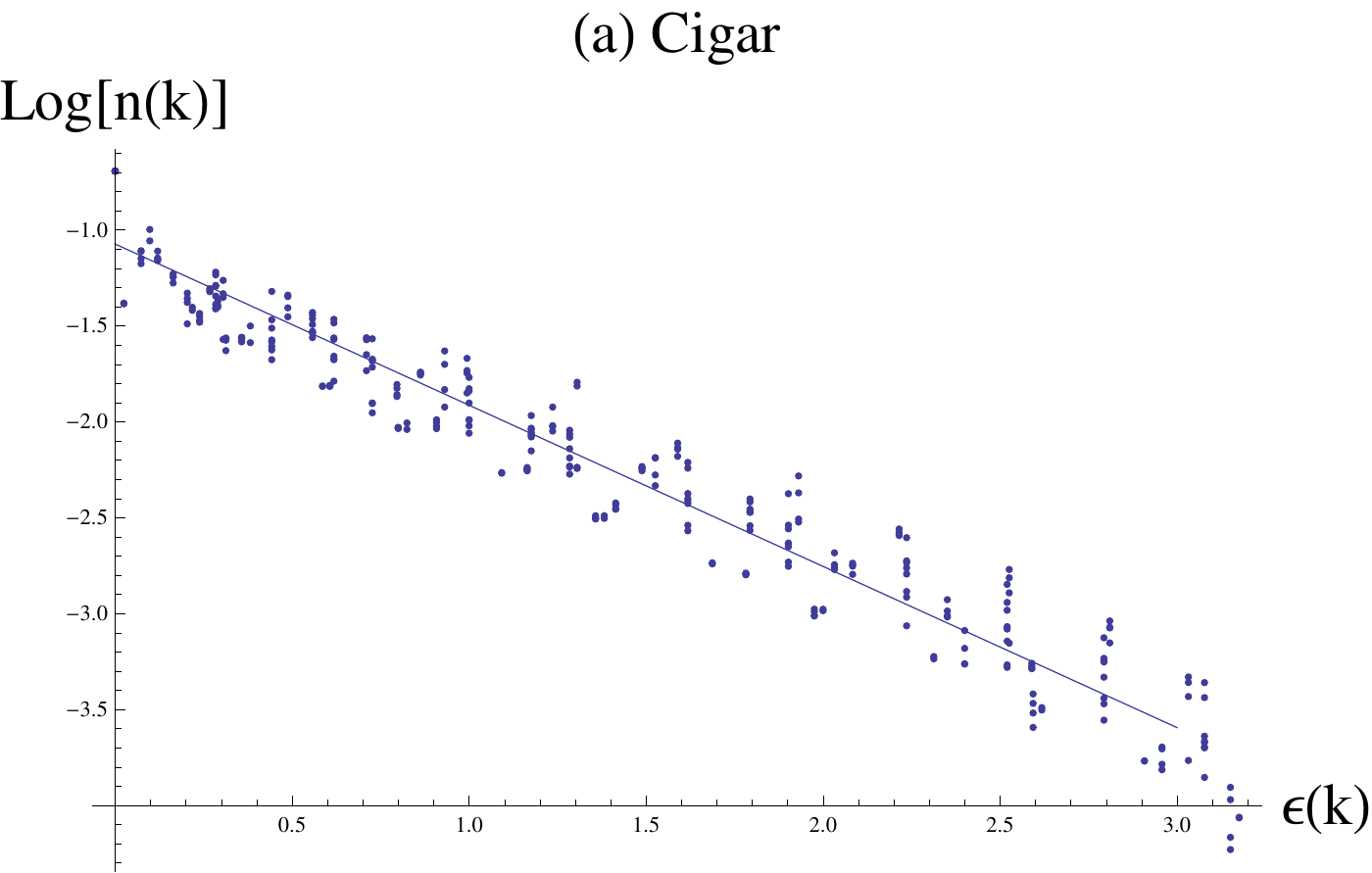}\\
\includegraphics[scale=0.5]{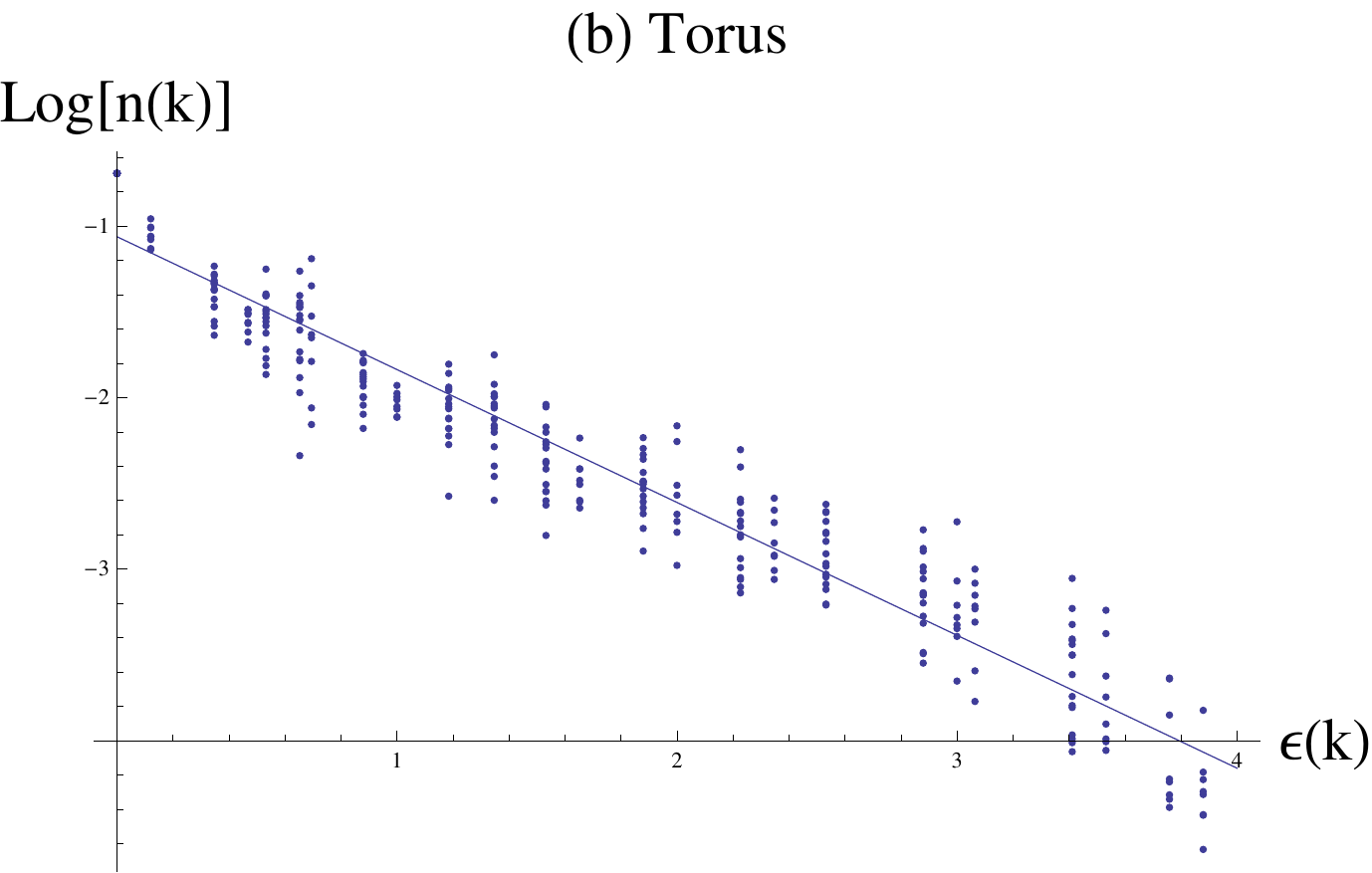}\\
\includegraphics[scale=0.5]{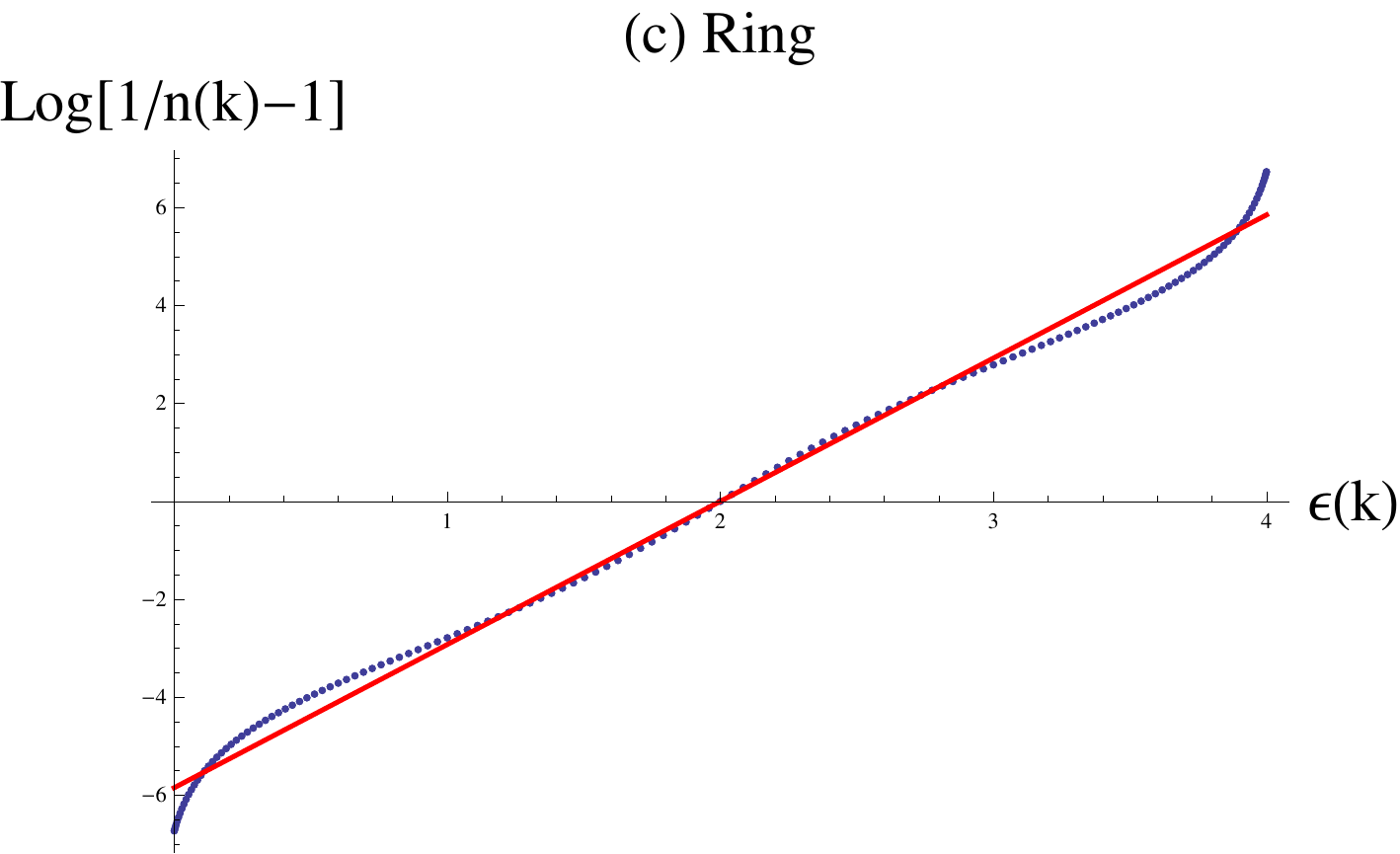}
\caption{Functions of the number of quasiparticles $n(k)=\langle \eta_k^\dagger \eta_k \rangle$, evaluated numerically, in the ground state of
the system with $\lambda=0.3$ for three different lattices: (a) cylinder [20x20], (b) torus [16x16] and (c) ring [300]. 
}
\label{fig:different-geometries}
\end{figure}

In Fig.~\ref{fig:different-geometries}, the populations obtained in the three geometries are plotted 
with respect to the energy for $\lambda=0.3$.
We observe that the observable is close to thermal in all cases. 
In particular, both for the cylinder and the torus, the energy modes as measured by a Unruh--de Witt detector are populated according to a Maxwell--Boltzmann distribution, 
\be
n_{MB}(k) \propto \e^{-\beta \epsilon(k)}\, ,
\ee
and therefore in the plan $(\log n(k), \epsilon(k))$ we see a straight line with slope $-\beta$.
For the case of a a circle, we observe a Fermi--Dirac distribution, 
\be
n_{FD}(k)=\frac{1}{\e^{\beta \epsilon(k)}+1}\, ,
\label{eq:Fermi--Dirac}
\ee
and the logarithm of the populations $\log(1/n(k)-1)$ scales linearly with $\epsilon(k)$.

For different values of the parameter $\lambda$ and size of the system, we numerically evaluated the Bogoliubov 
transformation between the modes and their populations $n(k)$.
In all cases, there are striking numerical evidences that:
\begin{itemize}
\item The temperature is well defined in the large lattice size limit (thermodynamic limit).
\item The temperature (inverse of the slope in the plots of Fig.~\ref{fig:different-geometries}) scales
	proportionally to  $\lambda$, \ie
	\be
		T\propto \lambda \, .\nonumber
	\ee

\item The higher the $\lambda$, the better is the alignment of the points on a straight line, that is, the 
        more thermal the
	behavior of the radiation. 
\end{itemize}

In order to better understand the thermal nature of these distributions, we solve in the next section the 1-dimensional chain with periodic boundary conditions 
analytically. 
 
%%%%%%%%%%%%%%%%%%%%%%%%%%%%%%%%%%
%%%%%%%%%%%%%%%%%%%%%%%%%%%%%%%%%%
%%%%%%%%%%%%%%%%%%%%%%%%%%%%%%%%%%

\subsection{Analytical solution of the 1-dimensional lattice}

In the one dimensional case, the Hamiltonian \eqref{eq:full-hamiltonian}
becomes
\be
H = -J\sum_{m=0}^{L-1}\left(a^{\dagger}_{m}a_{m+1}+\textrm{h. c.}   \right)
+ \lambda \sum_{m=0}^{L-1}\left(a_{m}a_{m+1} +\textrm{H. c.}  \right) ,
\label{eq:1d-model}
\ee
where $L$ is the length of lattice (chosen to be odd) and the periodic boundary conditions
are imposed by the identification of $a_0$ with $a_L$.
Let us point out that this Hamiltonian is equivalent to the XY model in an external magnetic field
\cite{fradkin}.

In order to diagonalize this model, let us exploit its translational invariance and introduce the Fourier modes
\begin{equation}
\eta_k=\frac{1}{\sqrt{L}}\sum_{m=0}^{L-1}\e^{-\iu \frac{2\pi k}{L}m}a_m\, .
\end{equation}
We note that according to this definition the mode $\eta_{-k}=\eta_{L-k}$.
From now on, we will use both $[-(L-1)/2,(L-1)/2]$
and $[0,L-1]$ as the range of $k$ depending on the simplicity of the notation.
In terms of these modes, the Hamiltonian \eqref{eq:1d-model} reads
\begin{eqnarray}
H&=&-J\sum_{k=0}^{L-1}2 \cos\left( \frac{2\pi}{L}k\right) \eta_k^\dagger \eta_k \nonumber \\
&+&\ \lambda\sum_{k=0}^{L-1}\left( \e^{-\iu \frac{2\pi}{L}k}\eta_k\eta_{L-k}+ h. c. \right) \, . \nonumber 
\end{eqnarray}

Reshuffling some terms and using the canonical anticommutation relations, the Hamiltonian can be written as 
\be
H=\sum_{k=0}^{L-1} 2 \epsilon(k) \eta_k^\dagger\eta_k + \iu  \sigma(k) (\eta_k \eta_{L-k} - \eta_{L-k}^\dagger \eta_{k}^\dagger)\, ,
\ee
where 
\begin{align}
\label{eq:dispersion-relation-1D}
\epsilon(k)&= -J \cos\left( \frac{2\pi}{L}k\right) \\ 
\sigma(k)  &= - \lambda \sin\left( \frac{2\pi}{L}k\right) \, .
\end{align}
Let us note that while $\epsilon(k)$ is a symmetric function $\epsilon(k)= \epsilon(L-k)$,  
$\sigma(k)$ is anti-symmetric, $\sigma(k) = -\sigma(L-k)$.

Thus, the Hamiltonian \eqref{eq:1d-model} is decomposed in $(L-1)/2$ independent Hamiltonians
\be
H=\sum_{k=0}^{L-1} H_k\, ,
\ee
with
\be
H_k= \epsilon(k) (\eta_k^\dagger\eta_k+\eta_{-k}^\dagger\eta_{-k})+\iu \sigma(k) (\eta_k^\dagger \eta_{-k}^\dagger-\eta_{-k} \eta_{k})\, ,
\ee
or, in matrix notation, 
\begin{equation}
\label{eq:matrix-form-Hk}
	H_k=	
	\left(
	\begin{array}{cc}
	\eta_k^\dagger &  \eta_{-k}
	\end{array}
	\right)
	\left(
	\begin{array}{cc}
	\epsilon(k) &  \iu \sigma(k) \\
	- \iu \sigma(k) & -\epsilon(k) 
	\end{array}
	\right)
	\left(
	\begin{array}{c}
	\eta_{k} \\ \eta_{-k}^\dagger
	\end{array}
	\right)
	.
\end{equation}
\subsubsection{Diagonalization}
First of all, let us note that
\begin{equation}
	\left(
	\begin{array}{cc}
	\epsilon(k) &  - \iu \sigma(k) \\
	\iu \sigma(k) & -\epsilon(k) 
	\end{array}
	\right)
=  \sigma(k) \sigma_y + \epsilon(k) \sigma_{z}\, ,
\end{equation}
where $\sigma_{y,z}$ are the standard Pauli matrices.
In the vector space formed by the $(1,\sigma_x,\sigma_y, \sigma_z)$, 
we have to rotate the vector $(0,0,\sigma,\epsilon)$ in order to get a vector of the form $(0,0,0,\omega)$.
Hence, one immediately realizes that the diagonalizing matrix is a rotation of an angle 
\be
\theta(k) = \arctan\left( \frac{\sigma(k)}{\epsilon(k)}\right)\, ,
\ee
around the X axis, 
\begin{equation}
U = \exp\left(\iu \frac{\theta}{2} \sigma_{x}\right) =  \cos(\theta/2)\mathbb{I} + \iu\sin(\theta/2)\sigma_{x} \, .
\label{eq:rotation-unitary}
\end{equation}
From Eqs.~\eqref{eq:matrix-form-Hk} and \eqref{eq:rotation-unitary},
each Hamiltonian $H_k$ can be diagonalized with a Bogoliubov transformation of the form:
\begin{equation}
\psi_{k} = \cos\left(\frac{\theta(k)}{2}\right) \eta_{k} - \iu \sin\left(\frac{\theta(k)}{2}\right) \eta^\dagger_{-k}\, .
\end{equation}
The Hamiltonian in terms of its eigenmodes reads
\be
H=\sum_k \omega(k) \psi_k^\dagger \psi_k \, ,
\ee
with
\be
\omega_k=\textrm{sign} (\epsilon_k) \sqrt{\epsilon^2(k)+\sigma^2(k)}\, .
\ee
The eigenvalues have then multiplicity two 
\begin{equation}
\omega(k) = \omega(L-k) =  \omega(-k)\, ,
\end{equation}
that is, the energy of a quasiparticle with momentum $k$ is the same as the energy a quasiparticle with momentum $-k$.

\begin{figure}
\includegraphics[scale=0.72]{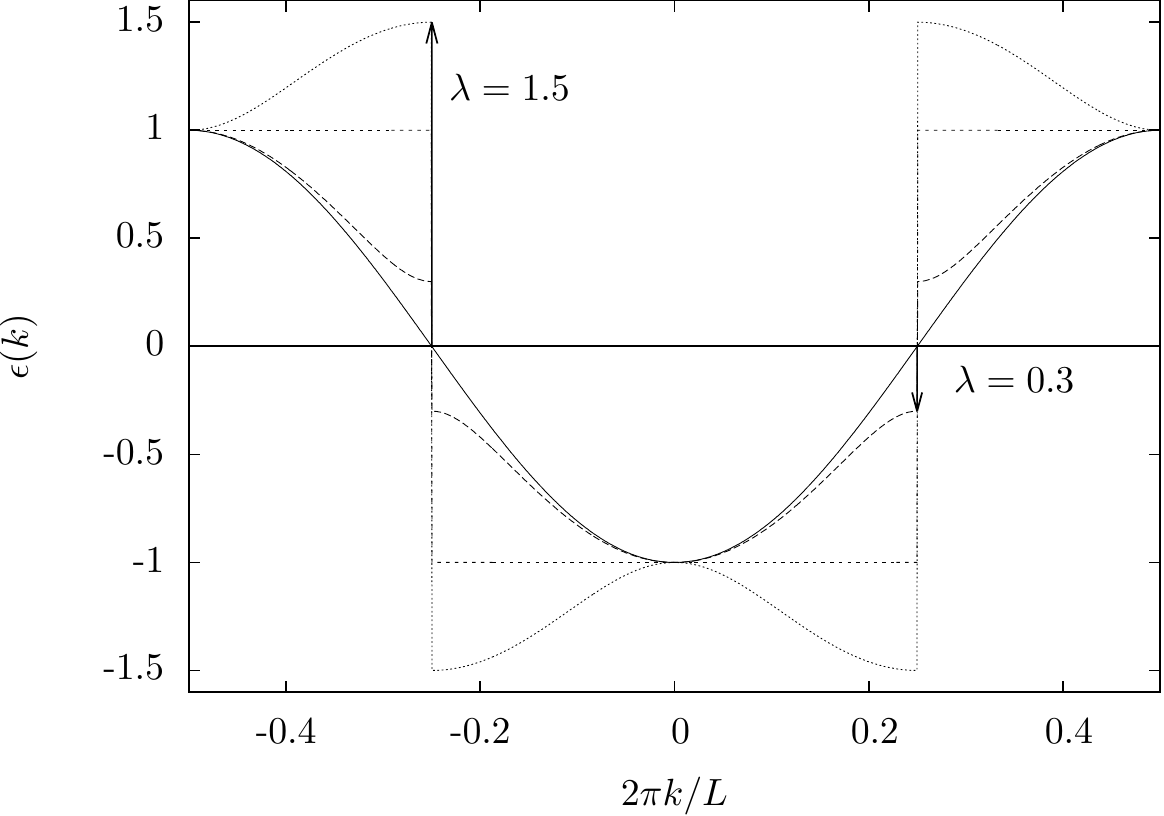}
\caption{Dispersion relation $\epsilon(k)$ for different values of $J$ and $\lambda$; $\lambda = 0$ (continuous), $|J|>|\lambda|$ (dashed),
$|J|=|\lambda|$ (square function), $|\lambda|>|J|$ (points).}
\end{figure}

\subsubsection{Ground state}
For simplicity, we fix $J=1$ and consider $\lambda > 0$ in units of $J$ by rescaling the Hamiltonian.
The ground state of the system with $\lambda=0$ is a Fermi sphere, and the first excited states are gapless modes 
($\epsilon(k)\propto (k-k_{F})$). In the limit of large $L$, this is a conductor.
 This is a direct consequence of the Goldstone theorem: the Fermi sphere 
ground state spontaneously breaks the global $U(1)$ symmetry of the original Hamiltonian, \ie
\begin{equation}
H(a,a^\dagger) = H(\e^{\iu \alpha}a,\e^{-\iu \alpha}a^\dagger)\, . \nonumber
\end{equation}

However, as expected from the fact that the $\lambda$
term breaks the global $U(1)$ symmetry of the standard action, the first excited states are separated
from the Fermi surface by a gap of $2\lambda$. In the limit of $L$ large, and large $\lambda$, this is an insulator. 
We can write then the ground state as
\begin{equation}
\ket{GS} = \prod_{|k| < k_F} \psi_{k}^{\dagger} \ket{0}_\psi\, ,
\label{eq:gs-def}
\end{equation}
where $\ket{0}_\psi$ is the Fock vacuum for the $\psi_{k}$ operators, and $k_F=(L-1)/4$ is the
the radius of the Fermi sphere ($\omega(k)<0$ for $|k|<k_F$). 

From Eq.~\eqref{eq:gs-def}, we can derive the structure of the ground state in terms of the operators $\eta_k$ and $\eta_k^\dagger$
(up to a global phase)
\begin{align}
\label{eq:initial-state}
\ket{GS}=&\prod_{k=0}^{k_F}(\sin(\theta(k)/2) - \iu \cos(\theta(k)/2) \eta^\dagger_{k}\eta^\dagger_{-k})\nonumber \\
&\times\prod_{k=k_F+1}^{(L-1)/2}(\cos(\theta(k)/2) + \iu \sin(\theta(k)/2) \eta^\dagger_{k}\eta^\dagger_{-k}) \ket{0}_\eta \, ,
\end{align}
and the occupation of the mode $k$ reads
\begin{equation}
n(k):=\bra{GS}\eta_k^\dagger\eta_k\ket{GS}=
\left\{ \begin{array}{ll}
\cos^2(\theta(k)/2) & \textrm{if $k\le k_F$}\\
\sin^2(\theta(k)/2) & \textrm{if $k > k_F$}
\end{array} \right. \, . \nonumber
\end{equation}
This piecewise function can be written in a more compact form by using the sine and cosine half angle formulas,
\begin{equation}
n(k)=\frac{1}{2}\frac{\sigma^2(k)}{\omega^2(k)+\epsilon(k)|\omega(k)|}\, .
\label{eq:radiation}
\end{equation}
Let us note that, fixed $\lambda$, the momentum distribution $n(k)$ 
only depends on its energy,
since $\sigma$ and $\omega$ can be expressed in terms of $\epsilon$ via
$\sigma^2=\lambda^2 (1-\epsilon^2)$ and $\omega^2=\sigma^2+\epsilon^2$.

This has a nice interpretation. Due to the presence of the Bogoliubov transformation,
the Fermi sphere appears to be slightly depopulated, with the particles being 
radiated
into the higher energy levels.

\subsubsection{Analogue thermal particle production}
In order to check whether the momentum distribution of the radiation measured by our detector in Eq.~\eqref{eq:radiation} is 
thermal, we have to compare it with the Fermi--Dirac distribution defined in Eq.~\eqref{eq:Fermi--Dirac}.

\begin{figure}
\begin{tabular}{l}
(a) \\ \includegraphics[scale=0.6]{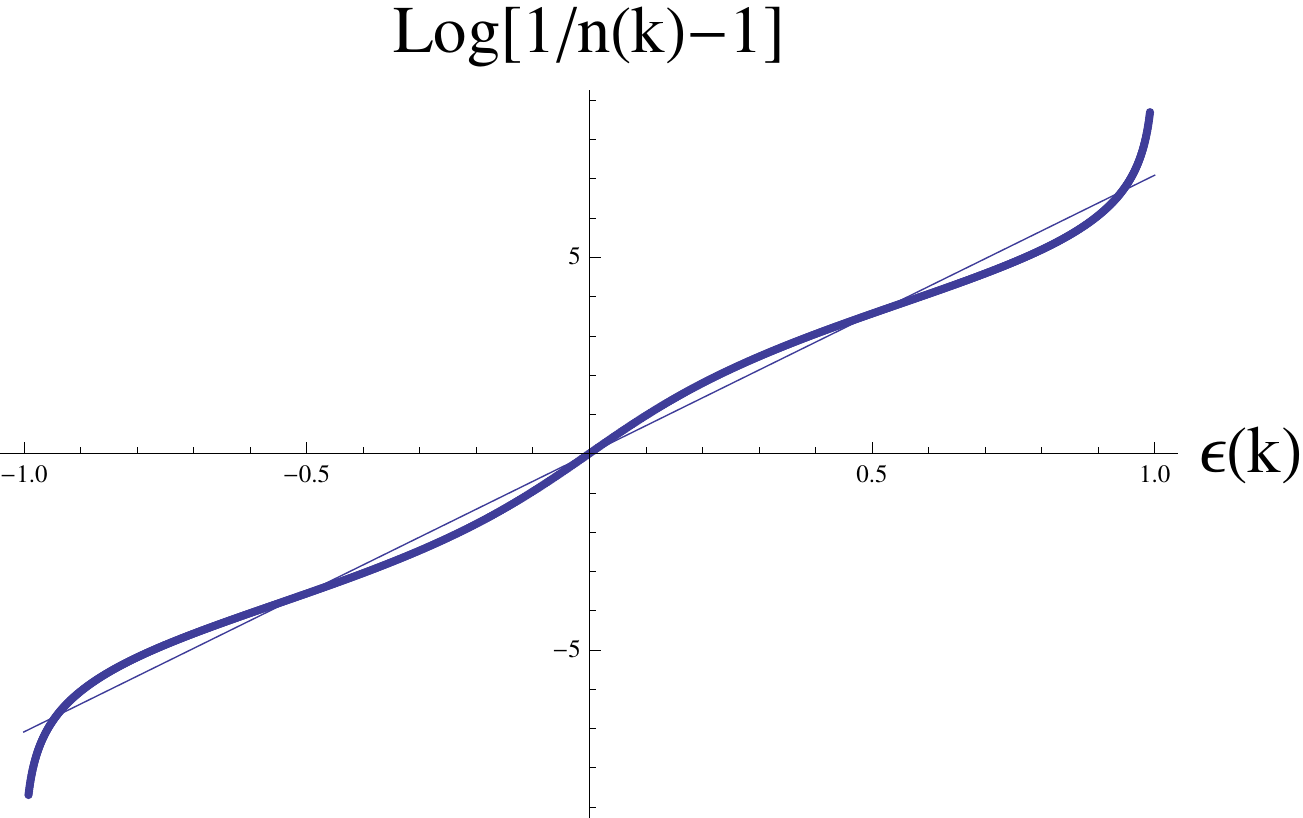}\\
(b) \\ \includegraphics[scale=0.6]{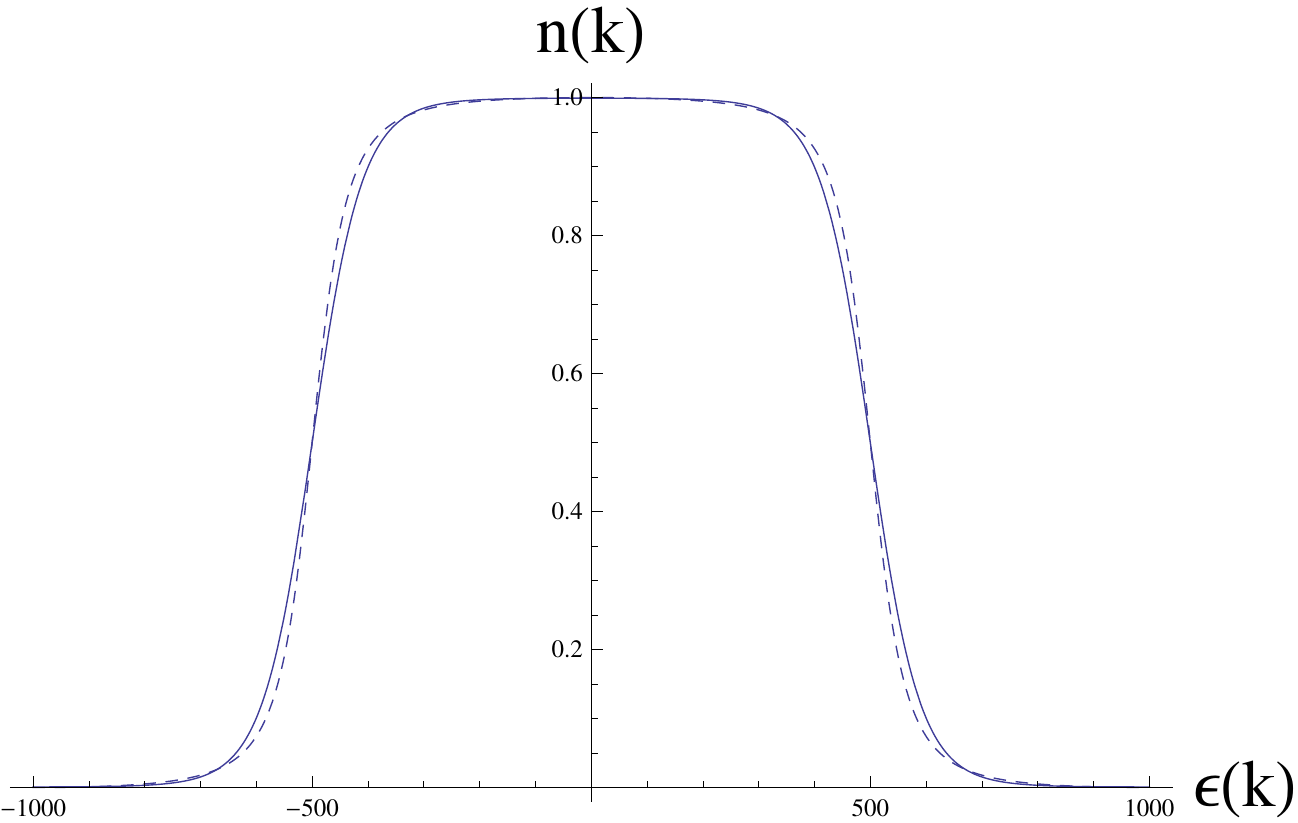}
\end{tabular}
\caption{ (a) Linear fit of the points $\left( \log(n(k)^{-1}-1),\epsilon(k) \right)$ 
where $n(k)$ is the momentum distribution of the modes $\eta_k$ in the ground state of the Hamiltonian \eqref{eq:1d-model}
and $\epsilon(k)$ is the energy of each mode. Its alignment implies that the occupations of the modes $k$ follow 
a Fermi--Dirac distribution with inverse temperature corresponding to the slope.
(b) Comparison between the real (continuous line) and the Fermi--Dirac (dashed line) distributions.}
\label{fig:temperature-fit}
\end{figure}

In Fig.~\ref{fig:temperature-fit}(a), the quantity $\log(n(k)^{-1}-1)$ is plotted with respect to $\epsilon(k)$
for several values of $\lambda$. If the radiation is thermal, these points should be fit by a straight line
that passes through the origin an has slope $\beta$.
This is precisely the behavior observed.
In Fig.~\ref{fig:temperature-fit}(b), the real population (continuous line) of each quasiparticle with momentum $k$ 
is compared to the population provided by the Fermi--Dirac distribution (dashed-line). 
Both distributions are very close to each other.

In order to get an analytical expression for the temperature of the radiation, 
let us notice that, if the distribution is a Fermi--Dirac, then
\begin{equation}
\beta = -4 \left. \frac{\dd n}{\dd \epsilon}\right|_{\epsilon=0} \, .
\label{eq:inverse-temperature-1D}
\end{equation}
In our case, the $n(k)$ distribution is very close to a Fermi--Dirac, and therefore,
the inverse temperature given by Eq.~\eqref{eq:inverse-temperature-1D} 
can be approximated from Eqs.~\eqref{eq:dispersion-relation-1D} and \eqref{eq:radiation},
\begin{equation}
\beta \simeq \frac{2}{\lambda},
\end{equation}
matching the expectation that $T \propto \lambda$.
%Let us emphasize that this relation between $\beta$ and $\lambda$ does not depend on the value of $\lambda$.
 
\subsection{Comment on Thermalization}
From the results of the last section we conclude that the momentum distribution of the particles that the detector would measure
is very close to thermal.
Let us note that, although the observables seem those of a thermal system, this does not mean that system is thermal.
There are other observables 
(apart from the number of particles $\eta_k^\dagger \eta_k$ considered previously)
for which the system will not look like thermal. 
But then, why does the momentum distribution $n(k)$ follow a thermal distribution?

In order to understand the origin of this thermal behavior, it is useful to write
the state of the system in terms of the eigenbasis of the Hamiltonian.
In Ref.~\cite{Riera2012}, thermalization of local observables
is explained by assuming a narrow energy distribution
together with typicality arguments \cite{Linden2009}.
In Ref.~\cite{Srednicki1994,Rigol2008}, thermalization is a consequence of a narrow energy distribution
and the Eigenstate Thermalization Hypothesis. 
In both cases, the temperature of the equilibrium state 
is determined by the position of the microcanonical energy window
in the spectrum of the Hamiltonian. 
Determining the energy distribution of the state could explain 
the approximated thermalization observed by means of one of these mechanisms.

Let us write the Hamiltonian \eqref{eq:H-detector} that defines our notion of particle in its spectral representation
for the 1-dimensional case case, 
\begin{equation}
\hdet=\sum_{k=0}^{L-1} \epsilon(k) \eta^\dagger_k \eta_k = \sum_{{\bf m}\in \{0,1\}^L} E_{{\bf m}}\proj{{\bf m}}\, ,
\label{eq:spectral-representation-H1D}
\end{equation}
where its eigenbasis $\{\ket{{\bf m}}\}$ is the Fock basis of the $k$ modes,
\begin{equation}
\ket{{\bf m}}=\prod_{k=0}^{L-1}(\eta_k^\dagger)^{m_k}\ket{0}_\eta\, ,
\end{equation}
with ${\bf m}$ the bit-string of occupations with components $m_k=\{0,1\}$ $\forall$ $k$.
The energy $E_{{\bf m}}$ of each eigenstate $\ket{{\bf m}}$ is given by
\begin{equation}
E_{{\bf m}}= \sum_{k=0}^{L-1} \epsilon(k) m_k \, .
\end{equation}
The first sum in Eq.~\eqref{eq:spectral-representation-H1D} runs over the $L$ different $k$-modes,
while the second one is performed over all the $2^L$ elements of the eigenbasis of the Hamiltonian.
The state of the system in this basis becomes:
\begin{equation}
\ket{GS}=\sum_{{\bf m}\in \{0,1\}^L} c_{{\bf m}} \ket{{\bf m}},
\end{equation}
with $c_{{\bf m}}=\langle {\bf m}\ket{GS}$.
These coefficents can then be derived from Eq.~\eqref{eq:initial-state} 
and written as
\begin{align}
c_{{\bf m}} &=\e^{\iu \varphi_{{\bf m}}} 
\prod_{k=1}^{k_F} \sin(\theta(k)/2)^{1-m_k}\cos(\theta(k)/2)^{m_k} \delta_{m_k,m_{-k}}\nonumber\\
&\times \prod_{k=k_F+1}^{(L-1)/2} \sin(\theta(k)/2)^{m_k}\cos(\theta(k)/2)^{1-m_k} \delta_{m_k,m_{-k}}\,   \, ,
\label{eq:pure-state-coefs}
\end{align}
where $\varphi_{{\bf m}}$ is a phase given by
\be
\varphi_{{\bf m}}=\frac{\pi}{2}\left(3\sum_{k=1}^{k_F}m_k+\sum_{k=k_F+1}^{(L-1)/2}m_k\right)\, .
\ee
The Kronecker delta $\delta_{m_k,m_{-k}}$ comes from the fact that $\ket{GS}$ is a superposition of 
only those Fock states with both modes $k$ and $-k$ either simultaniously occupied or empty (see Eq.~\eqref{eq:initial-state}).
This implies that only $2^{L/2}$ coefficents among the $2^{L}$ possible ones are non-trivial.

\begin{figure}
\includegraphics[scale=0.6]{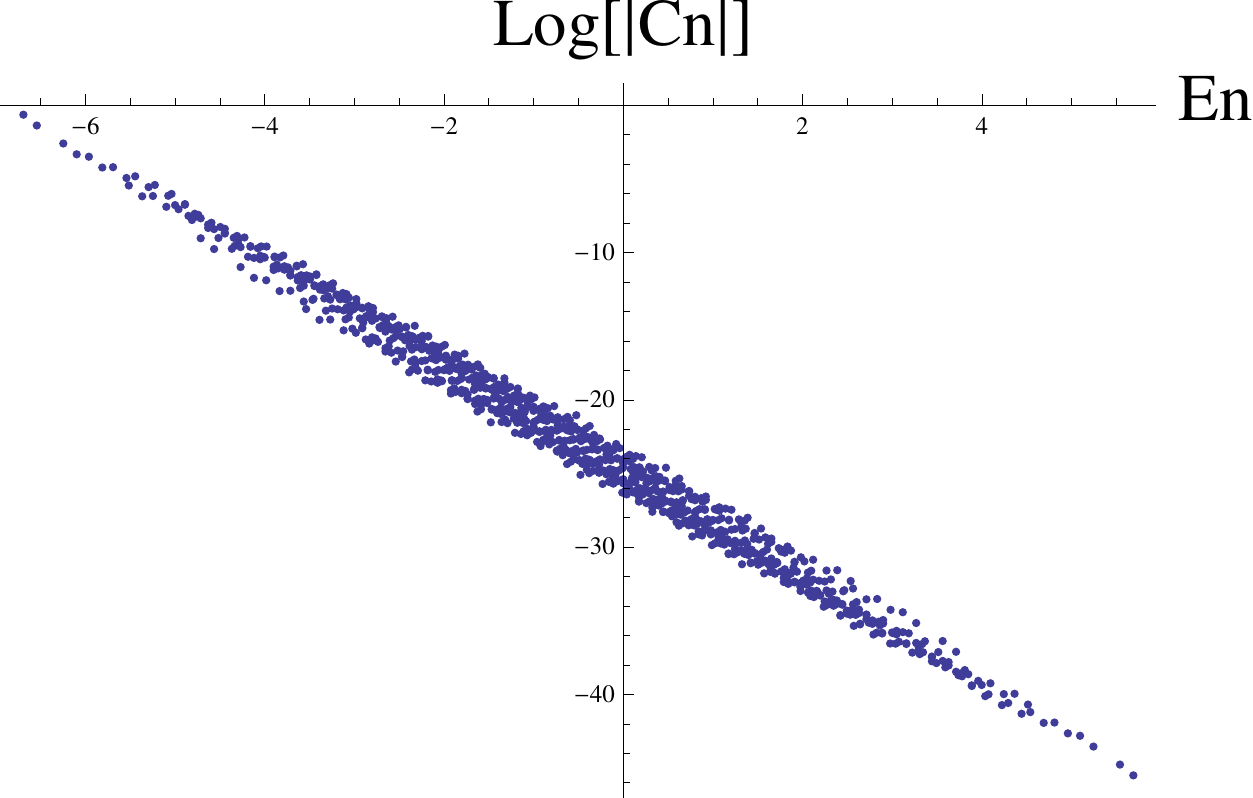}
\caption{Coefficients of the state of the system in the Hamiltonian eigenbasis with respect to the energy in a logscale in the vertical axis.
Their alignment in a straight line shows an exponential decay with the energy.}
\label{fig:purestatecoefs}
\end{figure}

In Fig.~\ref{fig:purestatecoefs}, the absolute value of the coefficients of the initial state
in the Hamiltonian eigenbasis
given by Eq.~\eqref{eq:pure-state-coefs} are plotted with respect to the energy. 
We observe that they decrease exponentially. The coefficient of the exponential decay can
be numerically determined, and we have to check that for any value of $\lambda$ it coincides with $\beta /2$.
The initial state can be then very well approximated by
\begin{equation}
\ket{GS}\simeq\sum_{{\bf m} \in M} \e^{\iu \varphi_{{\bf m}}}\e^{-\beta/2 E_{{\bf m}} }\ket{{\bf m}}\, ,
\label{eq:GSinHamEigenbasis}
\end{equation}
where the sum only runs over the set of bit-strings $M$ defined by
\be
M=\left\{{\bf m}\in \{0,1\}^L\, | \, m_k = m_{-k} \, \, \forall \, \, k  \right\}\, .
\ee

Any observable $\hat A$ which equilibrates, will equilibrate towards its time average expected value,
\be
\bar A = \lim_{\tau\to \infty} \frac{1}{\tau}\int_{0}^\tau \dd t \langle \hat A(t)\rangle = \trace (\hat A \omega)\, ,
\ee
where $\omega$ is the time average state defined by
\be
\omega=\lim_{\tau\to \infty} \frac{1}{\tau}\int_{0}^\tau\dd t \, \e^{-\iu \hdet t}\proj{GS}\e^{\iu \hdet t}\, .
\ee
Notice that $\omega$ is the apparent equilibrium state. We say ``apparent'' because the system is never in $\omega$, 
however, all those observables which equilibrate, do 
it towards the expected value that would be measured if the system was in $\omega$. 
Let us point out that in the case that the Hamiltonian has no degeneracies, 
$\omega$ is the completely dephased
state in the Hamiltonian eigenbasis and it maximizes the von Neumann entropy 
given the constants of motion \cite{Gogolin2011}.
Thus, the time average state $\omega$ reads in our case
\be
\omega=\sum_{{\bf m}\in M} \frac{\e^{-\beta E_{\bf m}}}{Z_M}\proj{{\bf m}}\, ,
\label{eq:omega-our-case}
\ee
with $Z_M=\sum_{{\bf m}\in M} \e^{-\beta E_{\bf m}}$. 

Now we can understand the origin of the Fermi--Dirac distribution
of the populations of each mode
\be
n(k)=\trace(\hat \eta^\dagger_k \eta_k \omega)=\sum_{{\bf m}\in \{M | m_k=1\}}\frac{\e^{-\beta E_{\bf m}}}{Z_M}
=\frac{1}{\e^{\beta \epsilon(k)}+1}\, ,
\nonumber
\ee
where we have used that 
$$\e^{-\beta E_{\bf m}}=\e^{-\beta \epsilon(k) m_k}\e^{-\beta \sum_{k'\ne k}\epsilon(k') m_{k'}}\, .$$

Let us remark that $\omega$ in Eq.~\eqref{eq:omega-our-case} is not a thermal state,
since the sum does not run over all the eigenstates of the Hamiltonian, but only
over those that have $m_k=m_{-k}$.
Nevertheless, in practice, all the observables that do not contain correlations
bewteen the modes $k$ and $-k$ will give the same value as if the system was in a Gibbs state.

In conclusion, we have seen that the thermal spectrum of the radiation
is a consequence of the exponential decay of the energy distribution,
that is, of the fact that the coefficients of the initial state
written in the Hamiltonian eigenbasis decay exponentially 
(see Fig.~\ref{fig:purestatecoefs} and Eq.~\eqref{eq:GSinHamEigenbasis}). 
This exponential decaying is not generic at all, in the sense that,  
most of Bogoliubov transformations relating the eigenbasis of two different Hamiltonians
(even if they have the same locality structure) will not produce it.
This represents another mechanism towards thermalization, where no bath is necessary
and where the temperature is not given by a macroscopic energy scale 
(the position of the microcanonical window of energies in the spectrum) but by the intensive
parameter which is quenched. 
It is an open question beyond the scope of this article to understand what properties 
two Hamiltonians have to share in order for the coefficents of an eigenstate of one Hamiltonian, 
spanned in the eigenbasis of the other, to decay exponentially respect to the energy.

%%%%%%%%%%%%%%%%%%%%%%%%%%%%%%%%%%
%%%%%%%%%%%%%%%%%%%%%%%%%%%%%%%%%%
%%%%%%%%%%%%%%%%%%%%%%%%%%%%%%%%%%
%%%%%%%%%%%%%%%%%%%%%%%%%%%%%%%%%%
%%%%%%%%%%%%%%%%%%%%%%%%%%%%%%%%%%
%%%%%%%%%%%%%%%%%%%%%%%%%%%%%%%%%%

\section{Application to trapping surfaces in Quantum Graphity}

% \subsection{Quantum Graphity intuition behi}

In this section we show how the model considered in the present paper can be derived from Quantum Graphity.
Moreover, we will apply the procedure to the trapping surfaces considered
previously in \cite{BlackOnions}.

\subsection{Derivation}
One of the most problematic features of Quantum Graphity models is their complexity. 
The dynamics of the graph makes the model hard to solve analytically. Thus,
approximate techniques to understand the behavior of the system are required.

First of all, we briefly describe the structure of the Hubbard model on a dynamical graph
introduced in Ref.~\cite{QGp2}. 
The Hilbert space of the system is $\mathcal H_\textrm{links} \otimes \mathcal H_{\textrm{particles}}$ with 
$\mathcal H_\textrm{links}=\bigotimes_{i=1}^{N_v(N_v-1)/2} \mathbb C^2$, $\mathcal H_{\textrm{particles}}= \bigotimes_{i=1}^{N_v} \mathcal H_i$
and $\mathcal H_i$ the local Hilbert space of one site. Its Hamiltonian can be written as
\begin{equation}\label{eq:full-model}
 H_{\textrm{total}} = H_{\textrm{links}}+  H_{\textrm{particles}} + H_{\textrm{BR}}\, ,
\end{equation}
where $H_{\textrm{links}}$ describes the dynamics of the graph (space),  $H_{\textrm{particles}}$ is the Hamiltonian for the
particles (matter), and $H_{\textrm{BR}}$ corresponds to the backreaction interaction between space and matter. More explicitly, these Hamiltonians read
\begin{align}
H_{\textrm{links}} &= - U \sum_{(i,j)} \sigma^z_{(i,j)}\, , \\
H_{\textrm{particles}} &= -t\sum_{(i,j)} P_{ij}\otimes  (a^\dagger_i a_j +a_i a^\dagger_j)\, , \\
H_{\textrm{BR}}&= \lambda \sum_{(i,j)}  \left(S^-_{(i,j)}\otimes  a^\dagger_i
a^\dagger_j +S^+_{(i,j)}\otimes a_i a_j\right),
\end{align}
where $P_{ij}=\proj{1}_{ij}$ is the projector onto the edge $(i,j)$, $S^+_{ij} = |1\rangle\langle0|_{ij}$ and $S^-_{ij} = |0\rangle\langle1|_{ij}$,
and $U$, $t$, and $\lambda$ set the speed of the oscillations of the links, the tunneling rate and the backreaction strength respectivelly.
Notice that in the Hamiltonian for the matter $H_{\textrm{particles}}$ the interaction among the particles has been neglected, but naturally it can be included. However, for the phenomena
we are interested in, we do not really need interactions, and we can work with free fields. 
First of all, let us assume a situation in which matter degrees of freedom do not affect the graph, i.e. their coupling to the
dynamical graph is small compared to hopping one. This means that the edges of the graph can be considered to be constant in time, and thus a fixed
graph in the first approximation. However, we will still retain the quantum structure of the graph, i.e. links can be still in a superposition of on and off states. 
In this approximation, we can derive the Hamiltonian (\ref{eq:full-hamiltonian}) from Quantum Graphity. 

In order to see this, we consider the matrix-elements of the Hamiltonian \eqref{eq:full-model} applied to a particular quantum graph, and realize that
these are the same as those of the Hamiltonian in Eq. \eqref{eq:full-hamiltonian}:
\be
\bra{\phi_m} H \ket{\phi_n} = \left(\bra{\phi_m}\otimes \bra{\psi_\textrm{space}}\right) H_{\textrm{total}} \left(\ket{\phi_n}\otimes \ket{\psi_\textrm{space}}\right) \, ,
\label{eq:eff-der}
\ee
with the quantum state for the links fixed to
\be
\ket{\psi_\textrm{space}}=\prod_{i,j} Q^{A_{ij}} \ket{0}^{\otimes \frac{N_v(N_v-1)}{2}}\, ,
\ee
where $Q$ is a Hadamard gate, $Q\ket{0}=1/\sqrt{2}(\ket{0}+\ket{1})$, and $A_{ij}$ is the adjacency matrix of the graph. 

The state $\ket{\psi_\textrm{space}}$ is then a product state of all the links in the state $\ket{0}$ except those links
that we would like to have active in the effective model, which are in the superposition
\be
\ket{+}=\frac{1}{\sqrt{2}}\left(\ket{0}+\ket{1}\right)=Q\ket{0} \, .
\ee
In order to prove Eq.\ \eqref{eq:eff-der} it is only necessary to realize that
\begin{align}
\bra{+}P\ket{+}&=\bra{+}S^\pm\ket{+}=\frac{1}{2} \\
\bra{0}P\ket{0}&=\bra{0}S^\pm\ket{0}=0 \, .
\end{align}
% These arguments show that the Hamiltonian used in the present paper can be derived from Quantum Graphity as an effective model.
We argue then that the Hamiltonian of Eq. \eqref{eq:full-hamiltonian} can be considered as an effective model in the case in which the graph is in a equal superposition
of on and off states.

Some comments are in order. First of all, let us notice that the bosonic model is unstable in the thermodynamic limit in which
the number of nodes of the graph goes to infinity. This is due to the fact that the energy can be decreased by increasing the number
of bosons in the graph. Morever, the model is reliable as an effective model of Quantum Graphity as
long as the number of particles $N_p< 2 L+N_i$, where $L$ is the number of links and $N_i$ is the number of particles in the states $\ket{\psi_{\textrm{total}}}=
\ket{\psi_{graph}}\otimes\ket{\psi_{part}}$.
The reason is that in Quantum Graphity the quantity $Q=N-2L$ is conserved. 

\subsection{Unruh--de Witt detectors}

In order to apply the procedure we described in the first part of the paper to Quantum Graphity, it is necessary to introduce the proper interpretation
to the Unruh--de Witt observables introduced previously. 
We use the fact that particle states are identified by the behavior of suitably defined
detectors, \ie devices that count the number of excitations associated to the measured
quantum state. As in the case of QFT in curved spaces, these detectors will be associated
to an observer, and hence an Hamiltonian with respect to which states are classified. 
In the case of QFT in curved spaces, particle production as the Unruh effect can be purely
kinematical phenomena associated to the mismatch between the notion of particle associated
to different observers, possessing, effectively, different Hamiltonians\footnote{The evolution
of an uniformly accelerated observer is controlled by the boost Hamiltonian.}.
Here we consider the ideal experiment of an observer that constructs his detector in a region
where the background graph is essentially regular (flat), and hence moves towards a
region of high connectivity, interpreted as a black hole analogue. The readings of these detectors
will be then influenced by the change in the notion of particle, ultimately associated to the
structure of the graph. The calculations, then, are the same as those in the first part of the paper,
although the quench involves also the structure of the graph, apart from the parameter $\lambda$.

\subsection{Difficulties with the particle production interpretation}
In \cite{Hamma,BlackOnions} it has been shown that certain graphs configurations might work effectively as black holes.
It has been shown that curvature is related to variations in the connectivity of the graph. 
A rather extreme case is
when the otherwise random graph contains a highly connected region, $\BB$, \ie a subset of nodes which
are connected by a link with almost any other node of the subset itself. In a certain limit (related to
the size of $\BB$) this region becomes a trapped region, \ie a region from which particles cannot
escape, once they enter it. 

As discussed in \cite{BlackOnions}, this phenomenon is rather peculiar with respect to 
the usual classical intuition in terms of bundles of geodesics. 
Indeed, it can be shown that the
lower energy eigenstates correspond to the case in which the particles are trapped into this region.
Moreover, the gap between the ground state and the excited states (with zero angular momentum) is proportional to the local degree (connectivity)
of the graph. The size of the
gap is finite for finite graphs, and hence these regions are not completely trapping. 
However, the larger is the size of this region (and thus the gap), the larger
energy needs to be transferred to a trapped particle to kick it out of it.

We are interested in the following question: can we see Hawking radiation? 
Unless we use this quench formalism, this is rather unlikely.

One way to see this is to say that, while the region $\BB$ behaves as a trapping region, it is not associated to an ergoregion
where negative energy modes can be stored. A more precise statement involves the 
unique well defined notion of ground state and particles so far available in these models. These are limitations typical of any ordinary non-relativistic
approach to this problem and as such, also Quantum Graphity suffers of it.

However, we will ask the following practical question. How will a particle detector react when
lowered near the region $\BB$? An important point to make explicit here is that the detector
will work as an external apparatus, \ie is not assumed to be built in terms of the same matter
fields living on the graph. If this black hole configuration were realized in a lab, the detector we
are referring to will be a scanner, a particle counter operated from the experimenter in the lab,
who would just count the clicks corresponding to suitably defined notion of particle. In turn
a particle is defined as an excitation above the ground state of a given Hamiltonian.

Formally, the setup described in the previous sections apply straightforwardly to the trapped surface lattice described above.
For increasing values of $\lambda$,
the logarithm of the population number becomes more and more a straight line with negative slope. This result is consistent with the findings in the 1-dimensional 
model. We thus expect that this phenomenon might be, to a certain extent, independent from the dimensionality of the lattice or from its connectivity.

We performed simulations for the lattice with cylindrical topology and with a complete graph attached to one end of the cylinder. This graph has been 
as a model of trapped surface in \cite{BlackOnions}. We report the results here for simplicity, as we do not believe the plots have particular relevance, other than
showing a particular slow convergence. The parameter $M$ which is the number of nodes in the angular coordinate of the cylinder, is also the size
of the complete graph. We observe that for increasing values of $M$, the temperature obtained from the fit of the logarithm of the population number is proportional to the 
value of $\lambda$. There is a reason for this, which we believe is easier to understand in the effective 1-dimensional model obtained in \cite{BlackOnions}.
In this case, the Hamiltonian for the modes with zero angular momentum is given, for $\lambda=0$:

\begin{equation}
 H_{\textrm{eff}}=- \sum_{i=0}^R J_i (a_i^\dagger a_{i+1} + \textrm{h. c.})
\end{equation}
where $J_0=M J$ and $J_i=J$ for $i>1$. When $M\gg 1$, the effective Hamiltonian of the model is the one with a node $0$ disconnected from the rest of the graph. In the original 2-dimensional model, this means that the complete graph 
can be considered as disconnected from the rest of the graph and the ground state becomes the product of the ground state for the complete region and the one for the cylindrical lattice, which then reduces the analysis to the one of the cylindrical
lattice. Thus, the complete graph, as a matter of fact, does not change the temperature of observed radiation but, indeed, the more the region is trapping, the more the temperature converges to the value
of the flat one. This is in fact a negative result, as it means that the radiation observed by the Unruh--de Witt detectors are not related to the trapping region, but only the quenching of the parameter $\lambda$.
%  This is also the reason why we believe that the presence of the trapped region does not contribute to the analog radiation in the quench formalism.

\section{Conclusions}

In this paper we investigated the possibility of using quantum quenches for simulating effects of quantum field theory in curved spacetime.
This research was inspired by ideas arised in the context of Quantum Graphity models, but can be placed also in the context of analogue models. 

A certain number of analogue models have been proposed in order to simulate gravitational phenomena or effects of quantum
field theory in the presence of curved geometries, \eg Hawking radiation and
cosmological particle production. In our case, our proposal is different from previous
models since we do not rely on the appearance of effective causal horizons, but rather
on a different origin of the mismatch between notion of particles as they are defined
for different detectors. 

In fact, we have shown that by considering Hamiltonians which do not conserve the particle number operator we can simulate approximate thermal particle production 
by means of a Quantum Quench. 
Quantum quenches are very interesting for many reasons. First of all, for many quantum mechanical systems it was shown that many observables
thermalize in the long time regime after a quantum quench. The mechanism is still poorly understood, even though it is thought that this might
be related to the eigenstates thermalization hypothesis \cite{Deutsch,Srednicki}. 
The setting we proposed is related, as we have seen, to the quantum quench of one parameter in the Hamiltonian.
We proposed to consider first the ground state of a 
Fermi--Hubbard model and then to perform a sudden quench of a parameter
associated to a particle number non-conserving term in the Hamiltonian.
We have shown analytically and numerically that the expectation value of Unruh--de Witt detectors is very close to thermal. This can be achieved by relating
the quenched and unquenched Hamiltonian through a Bogoliubov transformation. 

Some comments are in order. First of all, the expectation values of the number of particle in each mode do not evolve in time, thus their value just after the quench is the same they would have long after it.
This situation is similar to the one appearing in the context of quantum quenches, long after
the quench, 
with the important difference that,
contrarily to time-evolving observables, these can be measured right after the quench and do not need
to wait the ``relaxation'' of the system.

The search for thermality is a very common aspect in quench settings. 
While the total system is in a pure state, the expectation value of some observable are very close to thermal 
in the long time regime. 
This is confirmed by our calculations and numerical simulations, as the inspection of the occupation number of modes
corresponding to particles detected in analogue detectors. This could represent 
a new connection between the broad area of quantum systems in the laboratory and simulators
for gravity. 

However, the direct application to the case of black holes, as we have argued,
is not appropriate, given that the same particle creation effect is present in backgrounds that do not possess obvious causal boundaries. 

Indeed, when analyzing the case of
black hole-like graphs appearing in Quantum Graphity, to see whether they 
could be emitters of Hawking 
radiation, we could show that the temperature of the Planck distribution did not 
depend on the size of the trapped-surface, but indeed
converged to a value independent from the geometrical properties of the trapped surface. We gave an explanation for this,
based on the fact that the effective coupling between the trapping region and the rest of the graph decreases with the size
of the trapped surface.

Nonetheless, along the lines of \cite{Weinfurtner:2008if}, the setup that we described in the present paper could be used to simulate cosmological particle production, making the parameters of the Hamiltonian functions
of time, controlled by the experimenter.

\section*{Aknowledgements}
{\small
The authors are indebted to Alioscia Hamma for useful comments in the early stages of the work and Maurizio Fagotti for several comments. 
This research was supported by an NSERC and the Humboldt Foundation.  
Research at Perimeter Institute is supported by the Government of Canada through Industry Canada and by the Province of Ontario
through the Ministry of Research \& Innovation. F.C. is indebted to the University of Waterloo, the Perimeter Institute for Theoretical Physics and the 
Albert Einstein Institute in Potsdam for the hospitality and financial support 
during the period this work has been carried on.}

\bibliographystyle{unsrt}
\bibliography{biblio}

\end{document}